# Integration of 5G and Industrial Digital Models: A Case Study with AGVs

Jorge Cañete-Martín[1], Jorge Gómez-Jerez[1], M.Carmen Lucas-Estañ[1], Javier Gozalvez[1], Fernando Ubis[2]
[1]*Uwicore laboratory, Universidad Miguel Hernandez de Elche*, Elche (Alicante), Spain
[2]*Visual Components,* Espoo, Finland
{jcanete, jorge.gomezj, m.lucas, j.gozalvez}@umh.es, Fernando.Ubis@visualcomponents.com

***Abstract*—5G is a fundamental technology for the digitalization of smart manufacturing. Smart manufacturing relies on the use of digital models to optimize industrial processes before implementation on the manufacturing plants. These models should account for the impact of 5G communications to adequately dimension and optimize 5G-based industrial processes. This paper presents the first integration of industrial digital models with a 5G digital model, implemented as an Asset Administration Shell (AAS) of a 5G system. The two models are interconnected using an OPC UA-based interface. We evaluate the impact of the integrated model using a use case where Automated Guided Vehicles (AGVs) transport material from a warehouse to production lines. The AGVs periodically exchange their positions over 5G to avoid potential collisions. If the communications fail, the AGVs stop for safety reasons until a reliable 5G connection can be guaranteed. We demonstrate that, by integrating 5G and industrial digital models, it is possible to account for, and quantify, the impact of 5G communications on the operation and productivity of industrial processes. This result highlights the importance and necessity of integrating 5G into industrial digital models for their joint design and optimization.**

## I. Introduction

5G has been identified as a fundamental technology for the digitalization of smart and flexible manufacturing. 5G can meet the stringent communication requirements of industrial applications as well as provide seamless connectivity to mobile nodes, for example, Automated Guided Vehicles (AGVs). A wider adoption of 5G in manufacturing is somehow hindered to date by the need of solutions and tools that can facilitate the integration of 5G into industrial environments [1], including in the digital domain.

3D manufacturing software platforms are currently utilized to build digital models of industrial plants and optimize manufacturing processes in the digital domain before their physical deployment. These digital models reproduce the components (robots, conveyors, AGVs, sensors, actuators, and even workers) involved in the production processes, as well as their functions and interactions. The adoption of 5G in manufacturing requires the integration of 5G into the industrial digital models so that industrial processes can be optimized considering the impact that 5G communications can have on their operation. This is particularly relevant as 5G is not designed to sustain deterministic service levels, and the quality of service of 5G connections (e.g. reliability, latency, and throughput) can significantly vary when connecting mobile robots and AGVs.

The 5G Alliance for Connected Industries and Automation (5G-ACIA) already highlighted in [2] the need for developing digital models of 5G to facilitate its adoption and integration in manufacturing [3]. Following the recommendations in [2], the authors have developed a complete 5G digital model in [4]. The 5G digital model consists of an Asset Administration Shell (AAS) of the 5G User Equipment (UE) and an AAS of the 5G network (NW) that is fully aligned with 5G-ACIA guidelines as well as the Plattform Industrie 4.0 and 3GPP 5G standards. The study presented in this paper integrates the 5G digital model (implemented in Python) with a digital model of an industrial plant using OPC UA (Open Platform Communications Unified Architecture). The digital model of the industrial plant is implemented using Visual Components, a powerful 3D manufacturing software platform. We showcase the integration of the 5G and industrial digital models with a Mobile Robots use case where AGVs transport material from a warehouse to production lines. The AGVs periodically exchange their positions over the 5G network to avoid potential collisions. If the communications fail, the AGVs stop for safety reasons until a reliable 5G connection can be guaranteed. With this use case, we numerically quantify and demonstrate the impact that 5G communications (and its adequate deployment and planning) can have on the operation and productivity of industrial processes. These results demonstrate the importance and need of integrating 5G and industrial digital models to facilitate and foster a seamless adoption of 5G into smart manufacturing.

## II. Architecture and Interfaces

Fig. 1 presents the architecture designed to integrate the digital model of a 5G network and the digital model of an industrial production plant. The digital model of the production plant models the components that participate in the production processes (robots, AGVs, sensors, actuators, etc.) as well as their interactions. The 5G digital model consists of an AAS of the 5G Network (NW) and the AASs of the 5G User Equipments (UE). An AAS is a standardized digital representation of an asset that improves interoperability between heterogeneous components and systems by providing structured data models and standardized interfaces.

Each industrial device that requires transmitting data using 5G is connected to a UE in the 5G digital model. Within the 5G digital model, the UE interacts with the 5G NW to transmit the data and decide whether the data is correctly received and the QoS (Quality of Service) experienced on the 5G connection. The UE can report back this QoS information to the industrial device, e.g. to inform the device about which packets were successfully transmitted or the latency experienced on the 5G connection. With our integration of the 5G and industrial digital models, an industrial device can adapt its operation based on the quality of the 5G connection and the information received from the 5G digital model. For example, a production process may be halted or paused if its operation depends on receiving certain data through a 5G connection, and the transmission failed.

The proposed architecture is valid regardless of the specific software used for the implementation of the 5G and industrial digital models. In this study, we use the Visual Component software to build a digital model of an industrial production plant. The 5G digital model is implemented using the 5G AAS designed and implemented by the authors [5] using Basyx, and briefly presented in Section III. We utilize OPC UA to connect the industrial devices and 5G UEs; OPC UA is one of the technologies recommended by Plattform Industrie 4.0 [6] to implement AAS interfaces. We use a client-server communication where the industrial and 5G digital models act as clients and send requests to the OPC UA server. Visual Component provides a communication interface that enables direct communication between the industrial digital models and the OPC UA server. We implemented the necessary Application Programming Interfaces (API) so that our 5G digital model could interface with OPC UA and exchange information with the industrial digital model through the OPC UA server. These APIs have been implemented following the AAS interface specifications provided by Plattform Industrie 4.0 [6]. The interface defined in [6] is technology neutral, and HTTP/REST, MQTT, or OPC UA can be used for the implementation.

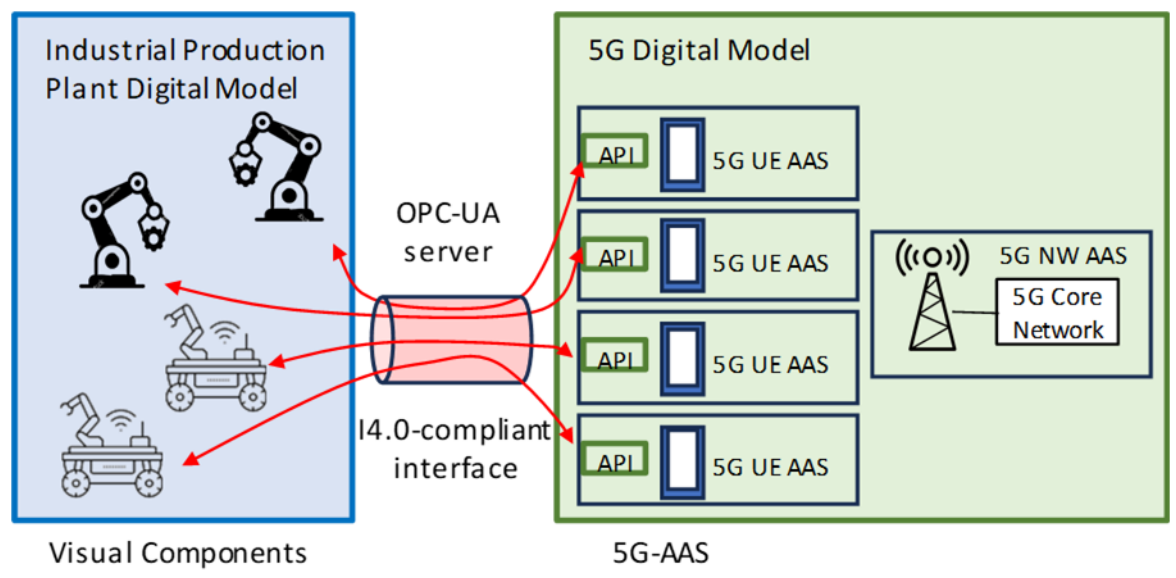

Fig. 1. Architecture for integrating 5G and industrial digital models.

## III. 5G Digital Model

The 5G digital model includes an AAS of the 5G UE and an AAS of the 5G NW. The UE serves as the termination point of a 5G link, and the NW models the most relevant functions of the 5G Radio and Core networks. The 5G AASs are defined in [4] and are openly available in [5]. They have been developed following the 5G-ACIA guidelines in [2] to facilitate the integration of 5G with industrial systems and applications, as well as the specifications of Plattform Industrie 4.0 for the definition of AASs [7] and 3GPP standards. The 5G UE and NW AASs have been implemented following a functional design where information is structured and grouped by functions or operations rather than by physical nodes. They both correspond to type 2 or reactive AASs following the classification from Plattform Industrie 4.0 in [7]. Type 2 re-active AASs utilize standardized interfaces (REST or OPC UA) to interact using client/server or publisher/subscriber models. The information provided by an AAS is organized in digital submodels [7], and the data contained in the submodels are referred to as submodel elements. Each submodel contains passive data and an active part. The passive data provides information about the characteristics of the AAS, and this information can be read or modified. The active part models certain technical functions or operations and has decision-making capabilities. Fig. 2 illustrates the structure and submodels of the 5G NW and UE AASs utilized in this study [4].

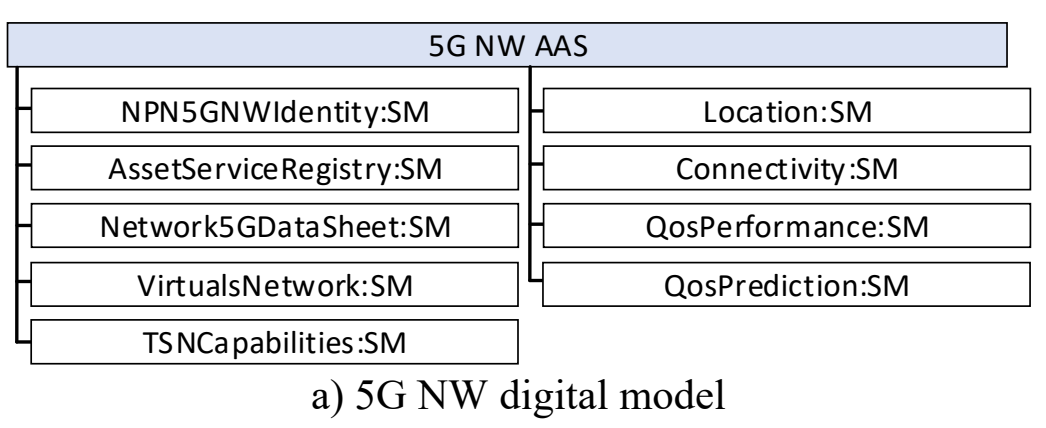

a) 5G NW digital model

5G UE AAS
Ue5GIdentification:SM
NetworkAccessRestrictions: SM
UE5GDatasheet:SM
Location:SM
UeAttachAndConnectionStatus:SM
QosMonitoring:SM

b) 5G UE digital model

Fig. 2. 5G AAS [4].

The 5G NW AAS includes submodels to identify the network, describes its characteristics at planning and deployment phases, and provide information about its technical capabilities (e.g. supported spectrum bands or transmission power) and the topology of the deployed 5G network. The 5G NW AAS also includes a submodel to describe the configured Virtual LANs (VLAN) and network slices, and a submodel with the parameters that characterize the 5G NW as a Time Sensitive Networking (TSN)-capable device to facilitate its integration with industrial TSN networks. 5G networks provide positioning services with significantly improved accuracy. In this context, the 5G NW AAS also includes a submodel containing information on the position of all connected UEs. The *Connectivity* submodel contains the list of active 5G connections and provides information about the UE and configured QoS profile for each active connection. This submodel also contains operations to establish/modify connections or the value of different RAN parameters. The *QoSPerformance* submodel provides information about the performance experienced in each active connection (e.g. latency, reliability, throughput). To this aim, it includes operations to estimate the performance experienced for a specific packet transmission or per active connection. The submodel also includes active operations that allow UEs or external (industrial) applications to monitor the value of specific performance parameters. Finally, the 5G NW AAS includes a submodel to expose data about QoS predictions.

The 5G UE AAS contains data about the UE identity, its technical characteristics, and information related to the physical and logical access restrictions of the UE to the 5G network. The 5G UE AAS also includes a submodel containing data about the location of the UE so that it can be made accessible to 5G-capable industrial devices. The *UEAttachAndConnectionStatus* submodel indicates the connection status of the UE and the list of active connections established by the UE. The *QosMonitoring* submodel contains information about the performance experienced by the UE, per connection or jointly considering all the active connections of the UE. The submodel also includes operations to subscribe to events from the 5G NW AAS that report about the performance experienced on a specific transmission or UE connection. It also includes operations so that 5G-capable industrial devices can subscribe to receive performance events' notifications.

## IV. Use Case

We demonstrate the integration of 5G digital models with industrial digital models, and its benefits, using the Mobile Robots use case presented in [8]. The use case considers AGVs that transport material from a warehouse to production lines in a pressing plant that transforms steel sheets for automobile door production. The pressing plant is depicted in Fig. 3, and is 57.86 m x 75 m. It is divided in three main areas: a warehouse where the steel sheets are stored before production, a production area with three parallel press lines, and a shipping warehouse where the final product is stored. The warehouse is separated from the production area by a concrete wall. Two AGVs carry the steel sheets from the warehouse to the press lines where they will be processed. AGVs periodically exchange over the 5G network their positions to avoid potential collisions among them. When an AGV does not receive the position of the other AGV for a specified period of time (referred to as survival time or $t_{sur}$), the AGV stops for safety reasons. The AGV will resume its movement after receiving the messages from the other AGV at the expected rate during a period $t_{res}$.

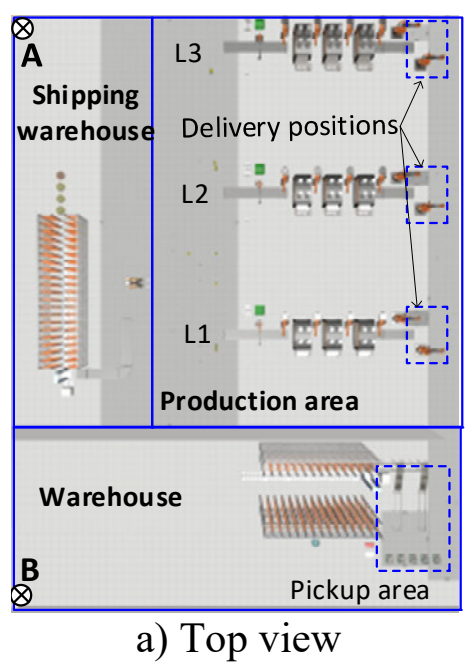

a) Top view

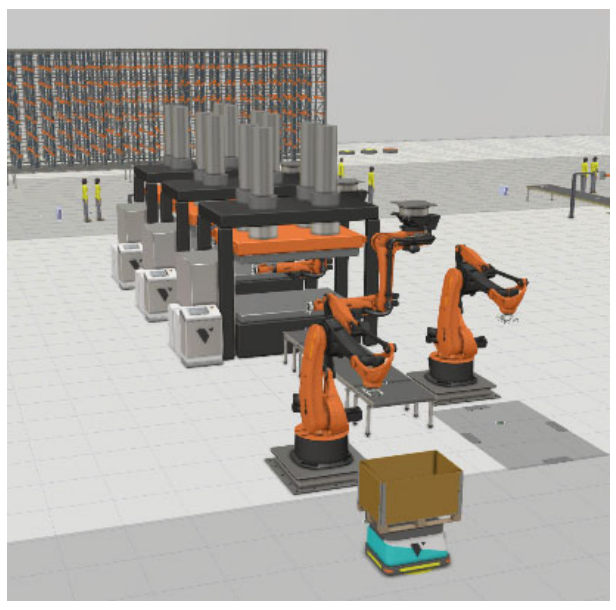

b) Press line and AGV.

Fig. 3. Industrial pressing plant.

Fig. 4 depicts the exchange of messages between the various components in the integrated 5G and industrial digital models. When AGV1 wants to send its position to AGV2, it sends the message to the 5G UE1 deployed on the AGV1. The 5G UE1 then informs the 5G NW that it has a new message to transmit. The 5G NW uses the *QosPerformance* submodel to model the 5G link performance and determine if the message transmissions from UE1 to the gNB and from the gNB to UE2 are satisfactory. The 5G NW notifies the 5G UE1 and UE2 whether the message has been transmitted with or without error. If the message is successfully transmitted on both links (UE1-gNB and gNB-UE2), UE2 notifies the AGV2 of the new position update received from AGV1.

## V. Experimental Setup

The industrial plant is modelled using Visual Components, a powerful software platform to digitally model industrial objects (machinery, robots, PLCs, etc.) and their interactions within production or industrial processes. The 5G digital model has been designed and implemented by the authors in Python using the open-source library Basyx Python SDK and the AASX Package Explorer. We used the asyncua library to implement the OPC UA server for interfacing the 5G and industrial digital models. The server requires a copy of the 5G UE and NW AASs in XML OPC UA Nodeset file format, which is obtained using AASX Package Explorer.

The plant has 5G coverage and we analyze two possible locations for the 5G base station or gNB: position A and position B in Fig. 3.a. We consider that AGV1 sends its position to AGV2 every second over the 5G gNB in order to coordinate their trajectories. Following [8], AGV2 stops to reduce the risk of collisions if it does not receive the position from AGV1 in a period $t_{sur}$ equal to 3 times the transmission period (which is equivalent to not receiving 3 consecutive messages). AGV2 resumes its movement when it receives 5 consecutive messages from AGV1 in 5 seconds. A message exchanged between the two AGVs is not delivered correctly

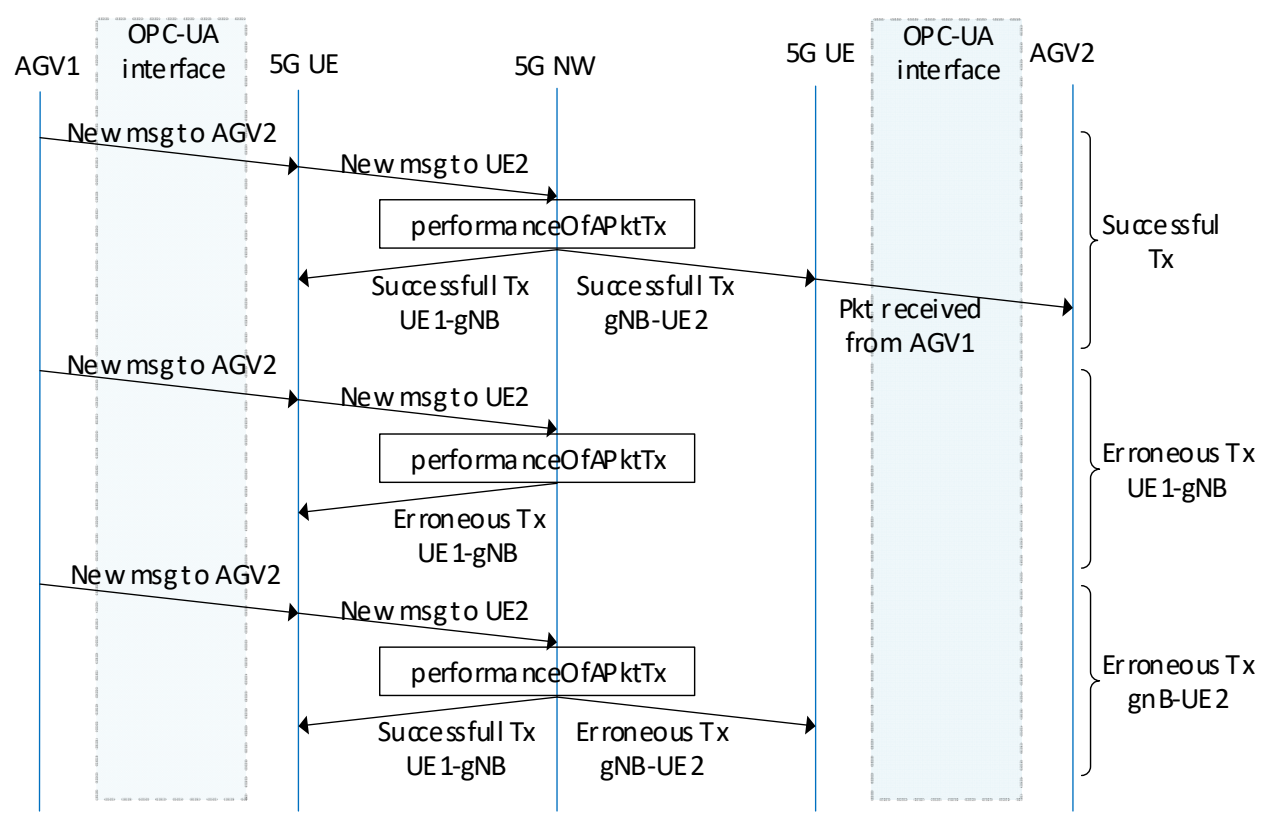

Fig. 5. Message exchange between different entities of the integrated 5G and industrial digital models.

if there is an error in the transmission from AGV1 to the gNB, or from the gNB to AGV2.

The 5G NW AAS models communication errors using Packet Delivery Ratio (PDR) curves that represent the probability of satisfactorily delivering a packet as a function of the distance between the transmitter and receiver. PDR curves are derived for various 5G modulation and coding schemes (MCSs) given their different error protection levels. We consider MCS10 that increases error protection at the cost of the spectral efficiency (2.57 bps/Hz), and MCS14 that increases spectral efficiency (3.61 bps/Hz) at the cost of a lower error protection. The PDR curves are derived using the model in [9] and Look-Up Tables (LUTs) obtained through link level simulations using the Matlab 5G toolbox [10]. The LUTs quantify the Block Error Rate (BLER) as a function of the experienced Signal to Noise Ratio (SNR) and the MCS. The LUTs are derived for the Indoor Factory channel (in particular, the TDL-C model) [11] and the 5G NR configuration for factory automation established in [12]. This configuration considers a 4GHz frequency carrier with 30 kHz subcarrier spacing. We assume a transmission power of 23 dBm and we consider non-line of sight propagation conditions and a 14dB attenuation from the concrete wall.

## VI. Numerical Evaluation

Fig. 5 shows an example of the messages transmitted with and without (w/o) errors on the links between AGV1 and the gNB (Fig. 5.a) and between the gNB and AGV2 (Fig. 5.b) over 25 minutes of simulation time. The results correspond to the scenario where the gNB is located in position A and the AGVs use MCS14 for their 5G transmissions. The quality of the transmissions depends on the distance between the AGVs and the gNB, but also on the presence or not of Line of Sight conditions between the AGVs and the gNB. Fig. 5 also identifies the time intervals during which AGV2 was stopped because there are 3 consecutive transmission errors.

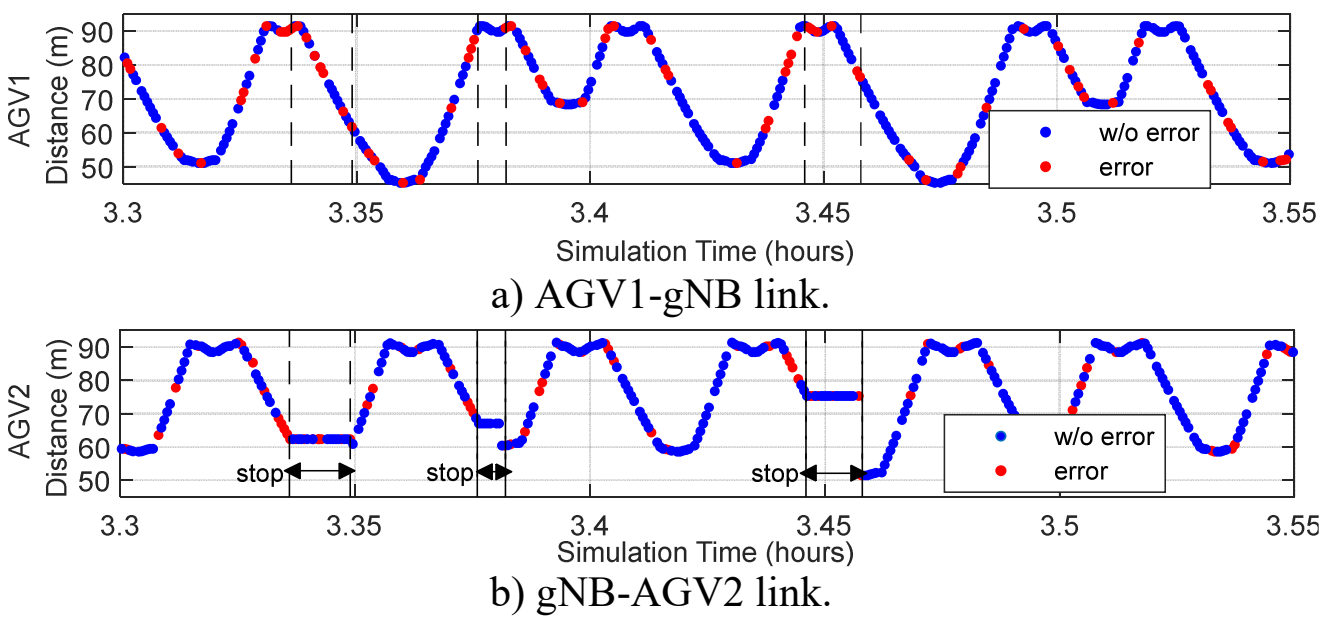

a) AGV1-gNB link.

b) gNB-AGV2 link.

Fig. 4. Example of the quality of 5G transmissions in the scenario.

Fig. 6.a shows the probability that AGV2 does not receive a message from AGV1 when using MCS10 or MCS14 for the 5G transmissions. This happens when there is a transmission error in either of the two links (AGV1-gNB and gNB-AGV2). The probability is depicted as a function of the distance of the AGVs to the gNB. The figure shows that the probability of message loss increases with the distance between the AGVs and the gNB, and the use of less robust MCS (i.e. MCS14). The highest probability of error occurs when both AGVs are at long distances to the gNB, which typically happens when the AGVs are in the pickup area if the gNB is in position A, or when both AGVs are delivering material to the production lines if the gNB is in position B. Fig. 6.b presents the normalized histogram of the distance of each AGV to the gNB throughout a simulation when the gNB is located in position A. The figure shows that the AGVs spend a significant percentage of time in the material pickup area (distances between 85 and 95 m). AGV1 is frequently located at distances of around 65 m and 45-50 m, which correspond to the delivery positions for lines 1 and 3. AGV2 spends a higher percentage of time at distances around 55-60 m, which correspond to the delivery position for line 2. This results in that 10.54% and 20.62% of the messages transmitted by AGV1 are not received at the AGV2 when using MCS10 and MCS14 respectively, as shown in Table I.

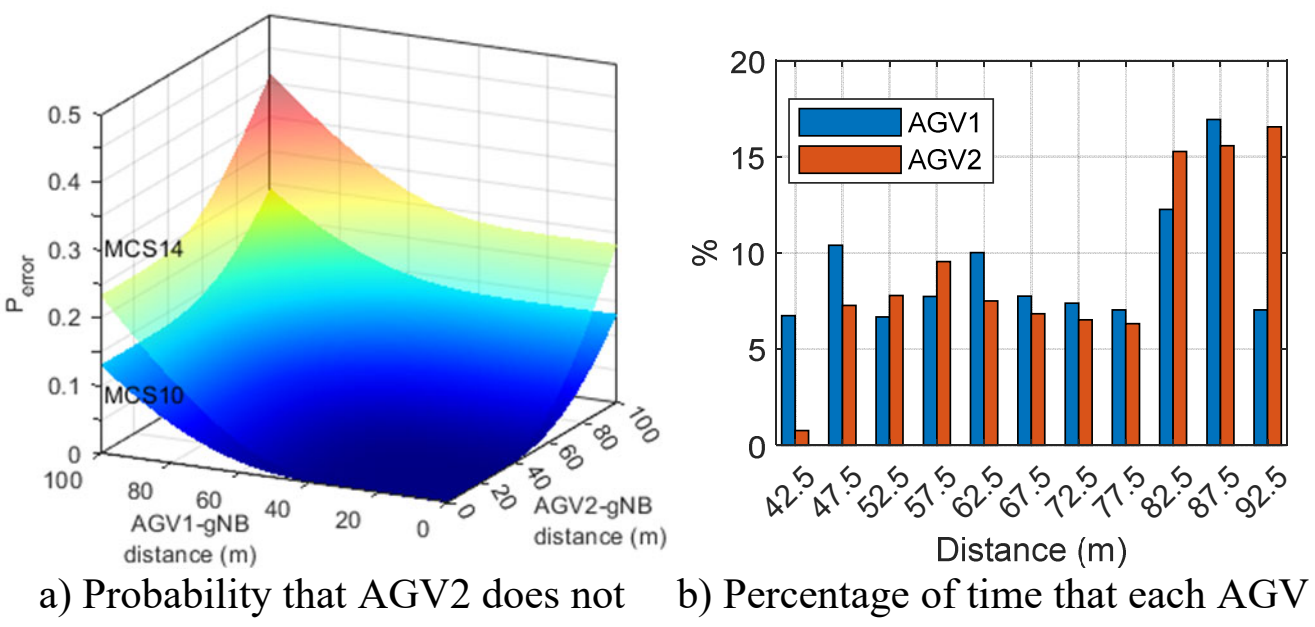

a) Probability that AGV2 does not receive a message from AGV1.

b) Percentage of time that each AGV is at a distance to the gNB.

Fig. 6. Communications and mobility statistics (gNB at position A).

Table I also shows the average time between consecutive stops of AGV2 (because it does not receive three consecutive location updates from AGV1), and the percentage of time that AGV2 is stopped during the simulation. The table shows that the average time between consecutive stops of AGV2 augments when using the less robust MCS (MCS14) because of the higher 5G transmission error rates (Fig. 6.a). This results in that AGV2 is stopped for 12.17% of the time when using MCS14 compared to just 1.24% when using MCS10. This is important because the time AGV2 is stopped impacts the production, as shown in Table II. Table II reports the reduction of produced items per press line and in all the plant considering the impact of 5G communications (with MCS14) compared to an ideal error-free scenario. Table II shows that the stops of AGV2 resulting from 5G communication errors mostly affect the third press line and reduces the number of produced items by 7.7% compared to an error-free scenario. Communication errors reduce the total production by 5.9%.

TABLE I. IMPACT OF 5G COMMUNICATIONS

| gNB position | MCS | % of messages not received at AGV2 | % of time AGV2 is stopped | Avg. time between stops (min) |
|---|---|---|---|---|
| A | 10 | 10.54% | 1.24% | 25.47 |
|  | 14 | 20.62% | 12.17% | 5.45 |
| B | 10 | 4.14% | 0.14% | 200 |
|  | 14 | 9.84% | 2.65% | 14.63 |

TABLE II. REDUCTION OF PRODUCED ITEMS COMPARED TO AN IDEAL COMMUNICATIONS ERROR-FREE SCENARIO

| gNB position | Press Line 1 | Press Line 2 | Press Line 3 | Total Production |
|---|---|---|---|---|
| A | 5.5% | 4.7% | 7.7% | 5.9% |
| B | 2.3% | 0.4% | 3.8% | 2.1% |

Table I and Table II report results also when the gNB is at position B. This deployment reduces the percentage of messages not received at AGV2 since the AVGs spend most of the time at the pickup area, and therefore at closer distances to the gNB compared to the scenario where the gNB is at position A. This augments the average time between consecutive stops of AGV2, reduces the percentage of time that AGV2 is stopped (Table I), and ultimately decreases the impact of 5G communication errors on the production of the press lines (Table II). The results in Table I and Table II show that the performance of 5G communications affect the industrial workflow, and an adequate deployment of 5G communications infrastructure is important to improve the productivity of industrial plants. These results also demonstrate the importance and need to integrate 5G and industrial digital models, and the value of the SW platform introduced in this paper for the efficient joint planning and dimensioning of industrial processes and 5G networks.

## VII. CONCLUSIONS

This paper has presented a first integration of industrial and 5G digital models using an OPC UA-based communication interface. We demonstrate the importance to integrate industrial and 5G digital models using a use case where AGVs transport material from a warehouse to production lines. The AGVs periodically exchange their positions over 5G to avoid potential collisions. If the communications fail, the AGVs stop for safety reasons until a reliable 5G connection can be guaranteed. Using the integrated digital model, we have analyzed how the deployment and configuration of the 5G network affects the operation of the AGVs and the production of the industrial plant. This result highlights the importance and necessity of integrating 5G into industrial digital models for their joint design and optimization.

## ACKNOWLEDGMENT

This work has been funded by European Union's Horizon Europe Research and Innovation programme under the Re4dy project (No 101058384), MCIN/AEI/10.13039/501100011033 and the “European Union NextGenerationEU /PRTR” (TED2021-130436B-I00), by the Spanish Ministry of Science and Innovation and Universities, AEI and FEDER funds (EQC2018-004288-P), and by Generalitat Valenciana (CIGE/2022/17), and UMH (VIPROAS23/11 and 2024).